\documentstyle[preprint,aps,epsfig]{revtex}
\tighten
\newcommand{\beq}{\begin{eqnarray}}
\newcommand{\eeq}{\end{eqnarray}}

\newcommand{\bpp}{B_d^0 \to \pi^+ \pi^-}
\newcommand{\spp}{S_{\pi\pi} }

\newcommand{\lpp}{\lambda_{\pi\pi} }
\newcommand{\app}{A_{\pi\pi} }

\newcommand{\rhob}{\bar{\rho} }
\newcommand{\etab}{\bar{\eta} }
\newcommand{\calb}{{\cal B} }
\newcommand{\calbb}{\overline{{\cal B}} }
\newcommand{\calo}{{\cal O} }

\newcommand{\ssb}{\sin{(2 \beta)} }

\newcommand{\tab}[1]{Table \ref{#1}}
\newcommand{\fig}[1]{Fig.\ref{#1}}

\newcommand{\im}{{\rm Im}\,}
\newcommand{\non}{\nonumber\\ }
\def \epjc{  Eur. Phys. J. C }

\def \npb{  Nucl. Phys. B }
\def \plb{  Phys. Lett. B }
\def \prd{  Phys. Rev. D }
\def \prl{  Phys. Rev. Lett.  }
\def \pr{   Phys. Rep. }

\begin{document}
\preprint{BIHEP-TH-2003-7}
\title{Constraints on the CKM angle $\alpha$ from the measured CP asymmetries
and branching ratios of $B \to \pi^+ \pi^-, \pi^+ \pi^0$ and $K^0\pi^+$ decays}
\author{Zhenjun Xiao\thanks{Email address: zjxiao@email.njnu.edu.cn} \\
{\small Department of Physics, Nanjing Normal University,
Nanjing, Jiangsu 210097, P.R.China} \\
Cai-Dian L\"u \thanks{Email address: lucd@ihep.ac.cn} \\
{\small CCAST (World Laboratory), P.O. Box 8730, Beijing 100080, P.R.China;}\\
and \\
{\small Institute of High Energy Physics,  P.O. Box 918(4), Beijing 100039, P.R.China
\thanks{Mailing address.}}\\
Libo Guo \\
{\small Department of Physics, Nanjing Normal University,
Nanjing, Jiangsu 210097, P.R.China} }
\date{\today}
\maketitle
\begin{abstract}
In this paper, we draw  constraints on the Cabibbo-Kobayashi-Maskawa (CKM) angle $\alpha$
and the strong phase $\delta$ from the experimental measurements of $\ssb$, the
CP-violating asymmetries and branching ratios of $B \to \pi^+ \pi^-, \pi^+ \pi^0$
and $K^0 \pi^+$ decays. In the SM, the measured $\ssb$ leads to an upper limit on
$\alpha$: $\alpha \leq 180^\circ-\beta$.
Taking the weighted-average of the newest BaBar and Belle measurements of $\spp$ and $\app$
in $B^0 \to \pi^+\pi^-$ decay  as the experimental input, we find the allowed range of
angle $\alpha$, $76^\circ \leq \alpha \leq 135^\circ$ for $r=|P/T| =0.3 \pm 0.1 $.
From the measured CP-averaged branching ratios of $B \to \pi^+ \pi^-, \pi^+
\pi^0$ and  $K^0 \pi^+$ decays, we found an inequality,
$\cos{\alpha} \cos{\delta} \geq 0.45$. The region of  $70^\circ \leq \alpha \leq 110^\circ$
is thus excluded by applying this inequality. The combined constraints on $\alpha$ and
$\delta$ are $117^\circ \leq \alpha \leq 135^\circ$ and
$-160^\circ \leq  \delta \leq -132^\circ$ if we take $\spp=0.49 \pm 0.27$ and
$\app=0.51 \pm 0.19$ as the experimental input. The new lower limit on the
angle $\alpha$ is dominated by the measured branching ratios considered in this paper,
and is much stronger than those obtained before.

\end{abstract}

%\vspace{2cm}

\newpage
\section{Introduction}

To measure the Cabbibo-Kobayashi-Maskawa (CKM) angles
$\alpha$, $\beta$ and $\gamma$ is one of the
main goals of the B  factory experiments. In the standard model (SM), the CP violation is
induced by the nonzero phase angle appeared in the CKM mixing matrix $V_{CKM}$, and the
unitarity of the matrix $V_{CKM}$ leads to the constraint
\beq
\alpha + \beta + \gamma = 180^\circ .\label{eq:pi}
\eeq

Recent measurements of $\sin{2\beta}$ in neutral B meson decay $B_d^0 \to J/\psi K_{S,L}^0$ by
BaBar \cite{87-091801,89-201802} and Belle \cite{87-091802,66-071102} Collaborations
established the third type CP violation (interference between the decay and mixing ) of
$B_d$ meson system. The world average of $\ssb$ dominated by the newest
BaBar and Belle measurements \cite{89-201802,66-071102} is \cite{0208080}
\beq
\ssb = 0.734 \pm 0.054,\label{eq:ssb}
\eeq
which leads to a twofold solution of the angle $\beta$:
\beq
\beta = \left ( 23.6^{+2.4}_{-2.2} \right )^\circ  \bigvee
\left ( 66.4^{+2.2}_{-2.4} \right )^\circ. \label{eq:betaexp}
\eeq
The first solution is in perfect agreement with the global fit results
as given for example in Refs.\cite{hocker01,he01,ckmfit}, while the second one
cannot be explained in the SM and requires the existence of new physics\cite{fm02}.
Several strategies to distinguish
these two solutions through a measurement of the sign of $\cos{2\beta}$ have been
proposed\cite{quinn00} recently. In this paper, we will use the first solution of the angle
$\beta$ as the experimental input.

After the experimental measurement of the angle $\beta$, more attention has turned to ways
of learning the angle $\alpha$ and $\gamma$. For the determination of  the angle $\gamma$,
the measured CP-averaged branching ratios of
$B \to \pi K, \pi  \pi$ decays play a key role, and have been studied intensively in the
literature \cite{prd57-2752,epjc6-451,pr370,xiao02a}. But
the direct constraint from experiments on $\gamma$ is still weak.

The CKM angle $\alpha$ ( or $\phi_2$ in Belle's word) can be determined by the experimental
measurements for the CP violating asymmetries of $\bpp$ decays\cite{fm02,gronau02,gr02,lx02}.
From the experimental measurements of
the branching ratios of $B \to \pi \pi$ and $K\pi$ decays, one can also extract out the CKM angle
$\alpha$\cite{gronau02,charles99,luo02}. Very recently,
based on the data sample of about $85$ and $88$ million $\Upsilon(4S) \to B\bar{B}$ decays,
BaBar and Belle collaboration reported their measurements of the CP violation of the $\bpp$
decay \cite{89-281802,0301032}, respectively.

\noindent BaBar result \cite{89-281802}:
\beq
\spp &=& 0.02 \pm 0.34 (stat.) \pm 0.05(syst.), \non
C_{\pi\pi} &=& -0.30 \pm 0.25 (stat.) \pm 0.04(syst.), \label{eq:babar}
\eeq
and  Belle \cite{0301032}:
\beq
\spp &=& -1.23\pm 0.41 (stat.) ^{+0.08}_{-0.07}(syst.), \non
A_{\pi\pi} &=& +0.77\pm 0.27 (stat.) \pm 0.08 (syst). \label{eq:belle}
\eeq

It is easy to see that the experimental measurements of the BaBar
and Belle collaboration are not fully consistent with each other: BaBar's result is still
consistent with zero, while the Bells's result strongly indicate nonzero $\spp$ and $\app$.
But we do believe that the current discrepancy between two collaborations will disappear
when more data become available. If we make a weighted average
\footnote{For the parameter $A_{\pi\pi}$ and $C_{\pi\pi}$, there
is a sign difference between the conventions of Belle and BaBar Collaboration:
 $A_{\pi\pi}=-C_{\pi\pi}$. We here use  Belle's convention\cite{89-071801}.} of the two
experiments, and find that
\beq
\spp &=& -0.49 \pm 0.27 (0.61), \\
\app &=& +0.51 \pm 0.19 (0.23), \label{eq:spp-exp}
\eeq
where the errors in brackets are those increased by the PDG scaling-factor
procedure \cite{pdg2002}.

In a previous paper\cite{lx02}, based on the data as reported by BaBar and Belle Collaboration
\cite{89-281802,89-071801}, we presented the general description of CP asymmetries of the
$B \to \pi^+ \pi^-$ decay and found the constraint on the CKM angle $\alpha$ and the
strong phase $\delta$.

In this paper, in order to refine the constraints on both $\alpha$ and $\delta$, we will
focus on the following experimental information: (a) the new measurements of CP-violating
asymmetries of $B\to \pi^+ \pi^-$ \cite{89-281802,0301032};
(b) the well measured CP-averaged branching ratios of
$B \to \pi^+\pi^-, \pi^+\pi^0$ and $K^0 \pi^+$ decays\cite{cleo-03,babar-03,belle-03};
and (c) the world average of $\ssb$ as given in Eq.(\ref{eq:ssb}) \cite{0208080}.
For the sake of completeness and comparison, we also quote the constraints on the CKM
angles from global fit \cite{hocker01,he01,ckmfit}.

This paper is organized as follows. In Sec.~\ref{sec-2}, we give a brief review about the
definition of the CKM angles, draw direct constraint on the
CKM angle $\alpha$ from the measured $\ssb$.
In Sec.~\ref{sec-3} we present the general description of CP asymmetries of the
$B \to \pi^+ \pi^-$ decay, consider new BaBar and Belle's measurements
of $\spp$ and $\app$ to draw the constraints on the CKM angle $\alpha$ and the
strong phase $\delta$. In Sec.~\ref{sec-4} we take the well measured
branching ratios of $B \to \pi^+ \pi^-, \pi^+ \pi^0$ and $K^0 \pi^+$ decay modes into account,
estimate the ratio of tree to penguin amplitude $|T/P|$, and finally draw
the constraints on both the CKM angle $\alpha$ and the strong phase $\delta$.
The conclusions are included in the final section.

\section{The CKM angles $\alpha$, $\beta$ and $\gamma$ in the SM} \label{sec-2}

In the SM with SU(2)$\times$U(1) as the gauge group, the quark mass eigenstates are not the
same as the weak eigenstates. The mixing between the down type quark  mass eigenstates was
described by a $3\times 3$ unitary matrix $V_{CKM}$ \cite{ckm}.
The elements of the CKM matrix $V_{CKM}$ are fixed by four parameters, one of which is an
irreducible complex phase.
Using the generalized Wolfenstein parametrization\cite{w183}, $V_{CKM}$ can be written as
\beq
V_{CKM}= \left(           \begin{array}{ccc}
          1-\frac{\lambda^2}{2} & \lambda & A \lambda^3 (\bar{\rho}-i \bar{\eta})\\
          -\lambda & 1-\frac{\lambda^2}{2} & A \lambda^2 \\
          A \lambda^3 (1-\rhob-i \etab )&-A \lambda^2 & 1
          \end{array} \right)   \label{eq:vckm}
\eeq
where $\rhob = \rho (1-\lambda^2/2)$, $\etab =\eta(1-\lambda^2/2)$, and $A$, $\lambda$, $\rhob$
and $\etab$ are the four independent Wolfenstein parameters. The parameter $A$ and
$\lambda$ have been measured with good precision,
\beq
\lambda=|V_{us}|=0.2196\pm 0.0026 \label{eq:ld-exp}
\eeq
from analysis of $K_{e3}$ decays\cite{pdg2002}, and
\beq
A=\frac{|V_{cb}|}{\lambda^2} = 0.854 \pm 0.042 \label{eq:a-exp}
\eeq
determined from the measured $|V_{cb}|=(41.2\pm 2.0)\times 10^{-3}$ and
$\lambda=0.2196\pm 0.0026$ \cite{pdg2002}.

\begin{figure}[hbt]%fig.1
\vspace{-120pt}
\begin{minipage}[]{\textwidth}
\centerline{\epsfxsize=1.2\textwidth \epsffile{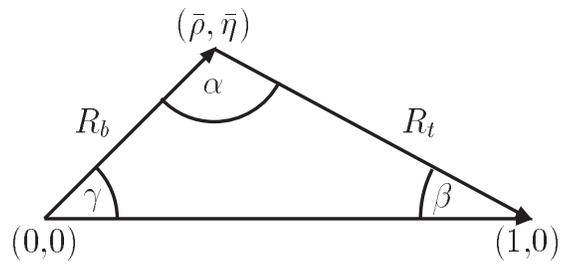}}
\vspace{-150pt} \caption{Unitarity triangle in $\bar \rho , \bar
\eta$ plane, corresponding to the $b \to d$ transition.} \label{fig:fig1}
\end{minipage}
\end{figure}

The unitarity of the CKM matrix implies  six ``unitarity
triangle''. One of them corresponding to the $b \to d$ transition yields
\beq
V_{ud}V_{ub}^* + V_{cd}V_{cb}^* + V_{td}V_{tb}^* =0. \label{eq:ut-1}
\eeq
This  unitarity triangle (UT) is just a geometrical presentation of this equation
in the complex $\rhob-\etab$ plane, as illustrated in Fig.\ref{fig:fig1}.
The three CKM angles in Fig.~\ref{fig:fig1} are defined as
 \beq \alpha &=&
 \arg\left(-\frac{V_{tb}^*V_{td}}{V_{ub}^*V_{ud}}\right), \label{eq:alpha}\\
 \beta &=&
    \arg\left(-\frac{V_{cb}^*V_{cd}}{V_{tb}^*V_{td}}\right),
  \label{eq:beta} \\
   \gamma &=& \arg\left(-\frac{V_{ub}^*V_{ud}}{V_{cb}^*V_{cd}}\right).
   \label{eq:gamma} \eeq
These definitions are independent of any kind of  parametrization of the CKM matrix
elements, and therefore universal. In the rescaled UT, $R_b$ and $R_t$ denote
the lengths of the two sides as shown in Fig.~\ref{fig:fig1}, and have been defined as
\beq
R_b & \equiv & \frac{|V_{ud}V_{ub}^*| }{ |V_{cd}V_{cb}^*|}=\sqrt{\rhob^2 + \etab^2}=
(1-\frac{\lambda^2}{2})\frac{1}{\lambda} \left | \frac{V_{ub}}{V_{cb}}\right |,
\label{eq:rb}\\
R_t & \equiv & \frac{| V_{td}V_{tb}^* |}{| V_{cd}V_{cb}^* |}=\sqrt{(1-\rhob)^2 + \etab^2}=
\frac{1}{\lambda} \left | \frac{V_{td}}{V_{cb}}\right |.
\label{eq:rt}
\eeq

In terms of $(\rhob, \etab)$, $\sin(2\phi_i)$ ($\phi_i=\alpha,
 \beta, \gamma$) can be written as
 \beq
 \sin(2\alpha)&=&\frac{2\etab(\etab^2+\rhob^2-\rhob)}{(\rhob^2+\etab^2)((1-\rhob)^2
   +\etab^2)},\label{eq:sin2a}\\
 \sin(2\beta)&=&\frac{2\etab(1-\rhob)}{(1-\bar\varrho)^2 +  \etab^2},  \label{eq:sin2b}\\
 \sin(2\gamma)&=&\frac{2\rhob \etab}{\rhob^2+\etab^2}.
 \eeq

Within the framework of the SM, intensive studies have been done to constrain the UT
by a global fit using the currently available data on neutral K, B mixing, semi-leptonic
B and K meson decays\cite{hocker01,he01,ckmfit},
such as $|V_{us}|, |V_{cb}|, |V_{ub}|$, $\epsilon_K$, $\Delta M_d$ , $\Delta M_d/\Delta M_s$,
and $\ssb$. The constraint in the $\rhob-\etab$ plane resulting from Eq.~(\ref{eq:rb}),
for example, is represented by a circle of radius $R_b$ that is centered at
$(\rhob,\etab)=(0,0)$, while the measured $\epsilon_k$ may fix a hyperbola in the
$\rhob-\etab$ plane. The commonly allowed region for the apex of the UT in the $\rhob,\etab$
plane corresponds to the so-called global fit result.
The ranges for the CKM angles $\alpha, \beta$, and $\gamma$
obtained by a recent ``CKM fit" are \cite{hocker01}
\beq
80^\circ \lesssim \alpha \lesssim  126^\circ, \ \
14^\circ \lesssim \beta \lesssim 27^\circ, \ \
34^\circ \lesssim \gamma \lesssim 82^\circ. \label{eq:alpha-1}
\eeq
And the new global fit results can be found in Ref.\cite{ckmfit}, the typical ranges
for the CKM angles are \cite{fleischer03}
\beq
70^\circ \lesssim \alpha \lesssim  130^\circ, \ \
20^\circ \lesssim \beta \lesssim 30^\circ, \ \
50^\circ \lesssim \gamma \lesssim 70^\circ. \label{eq:alpha-2}
\eeq
It is easy to see that the indirect constraint on the angle
$\beta$ agrees very well with the first solution
$\beta = (23.6^{+2.4}_{-2.2})^\circ$ in Eq.(\ref{eq:betaexp}) from the BaBar and Belle's
measurements, but strongly disfavor  the
second solution $\beta \approx 66^\circ$. We therefore take $\beta \approx 23.6^\circ$ as the
physical solution in our analysis and treat the limit
\beq
\alpha \leq 180^\circ -\beta = 156.4^\circ\label{eq:al-1}
\eeq
as the direct physical upper limit on the angle $\alpha$.

It is worth to mention that the constraints on the CKM angles from global analysis
also depend on the specific statistical method used in the analysis (for example, the
Bayesian, Rfit, or Scanning method, etc. ), but the resulted difference is not
significant. For more details about the global fit of the UT parameters, one
can see the new CERN Report \cite{ckmfit} and references therein.

\section{CP asymmetries of $B \to \pi^+ \pi^-$ decay} \label{sec-3}

\begin{figure}[hbt]
\vspace{-120pt}
\begin{minipage}[]{\textwidth}
\centerline{\epsfxsize=1.2\textwidth \epsffile{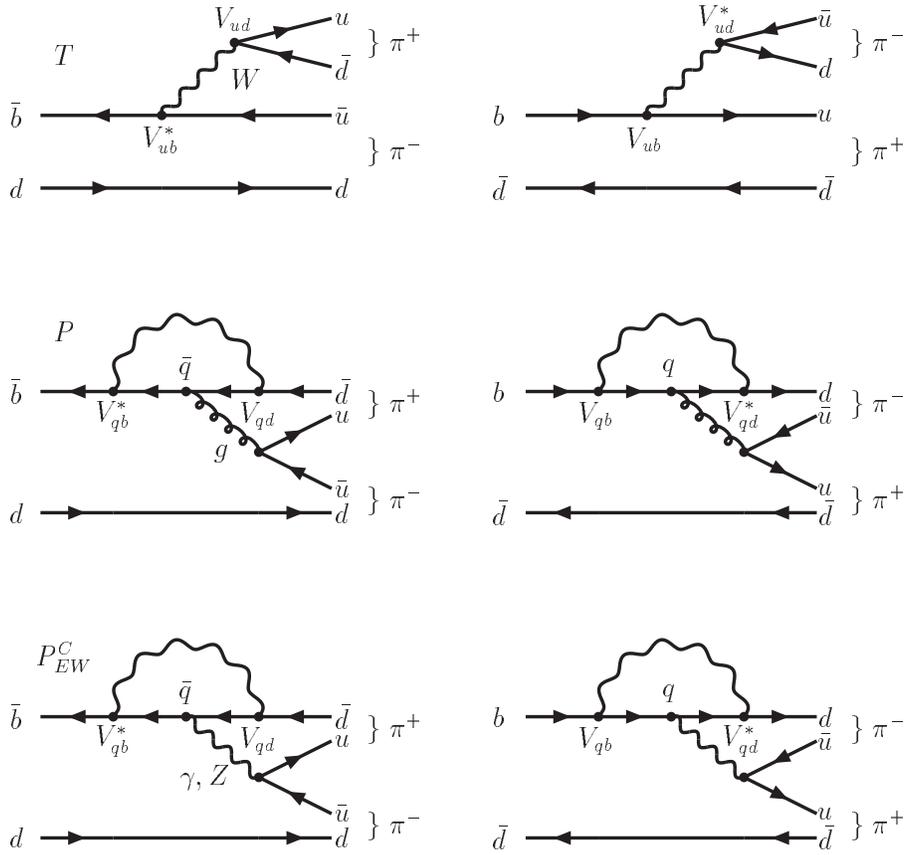}}
\vspace{-150pt}
\caption{The tree (T),  QCD penguin (P) and color-suppressed electroweak penguin
($P_{EW}^C$) diagrams with $q=u,c,t$,  contributing  to the
$B_d^0, \bar{B}^0_d \to \pi^+ \pi^-$ decays. }
\label{fig:fig2}
\end{minipage}
\end{figure}

The $B \to \pi^+ \pi^-$ decay mode plays an important rule in measuring the angle $\alpha$.
As shown in \fig{fig:fig2}, both the tree and penguin diagrams contribute to the
$B \to \pi^+ \pi^-$ decay simultaneously. The tree (T) contribution comes from $b \to u\bar{u}d$
transition by exchanging a W boson, while there are QCD penguin (P) and
color-suppressed electroweak penguin ($P_{EW}^C$) contributions with internal
quarks $q=(u,c,t)$. The suppressed annihilation diagrams have been neglected.

\subsection{Parametrization of $B \to \pi^+ \pi^-$ decay} \label{sec-31}

From Fig.\ref{fig:fig2}, the decay amplitude $A(B_d^0 \to \pi^+ \pi^-)$ and
its conjugate $\bar{A}(\bar{B}_d^0 \to \pi^+ \pi^-)$ can be written as \cite{sanda}
\beq
A &\equiv & A(B^0 \to \pi^+ \pi^- ) =-\left [  V_{ub}^*V_{ud} T
    + V_{tb}^*V_{td} P_t + V_{cb}^*V_{cd} P_c + V_{ub}^*V_{ud} P_u \right ] ,  \\
\bar{A} &\equiv & \bar{A}(\bar{B}^0 \to \pi^+ \pi^- )= - \left [ V_{ub}V_{ud}^* T
    + V_{tb}V_{td}^* P_t + V_{cb}V_{cd}^* P_c + V_{ub}V_{ud}^* P_u \right ],
\eeq
where $T$ is the tree amplitude, and $P_i$ $(i=u,c,t)$ are penguin amplitudes.
Using the CKM unitarity relation as given in Eq.(\ref{eq:ut-1}), the decay amplitude $A$ can be
rewritten in three different ways
\footnote{This is the so-called CKM ambiguity \cite{lss99}.}
\cite{ty03}
\beq
{\rm Convention\ \ A:}\ \  && A = -\left [ V_{ub}^*V_{ud}  T_{uc} + V_{tb}^*V_{td} P_{tc}\right ] ,
 \label{eq:a1a} \\
{\rm Convention \ \ B:}\ \  && A= -\left [ V_{ub}^*V_{ud}  T_{ut} - V_{cb}^*V_{cd} P_{tc}\right ] ,
 \label{eq:a1b} \\
{\rm Convention \ \ C:}\ \  && A= -\left [- V_{tb}^*V_{td}  T_{ut} -  V_{cb}^*V_{cd} T_{uc}\right ] ,
 \label{eq:a1c}
\eeq
with
\beq
&& T_{ut} \equiv  T + P_u - P_t, \ \  T_{uc} \equiv  T + P_u - P_c, \\
&& P_{tc}  \equiv P_t -P_c .
\eeq
The first two conventions are frequently used in previous studies\cite{gronau02,gr02,lx02,luo02},
the third one is also considered in Ref.\cite{ty03}.

In our analysis, we use Convention A and write the amplitudes $A$ and $\bar{A}$
in the following way
\beq
A&=& -\left [ V_{ub}^*V_{ud}  T_{uc} + V_{tb}^*V_{td} P_{tc}\right ]=
 - \left [ e^{i\gamma} e^{i\delta_T} |T| + e^{-i\beta} e^{i\delta_P} |P| \right ] ,
\label{eq:a11} \\
\bar{A}&=& -\left [ V_{ub}V_{ud}^*  T_{uc} + V_{tb}V_{td}^* P_{tc}\right ]=
- \left [ e^{-i\gamma} e^{i\delta_T} |T| + e^{i\beta} e^{i\delta_P} |P| \right ].
\label{eq:a21}
\eeq

The SM  predicts  the CP-violating asymmetries in the
time-dependent rates for initial $B^0$ and $\bar{B}^0$ decays to a
common CP eigenstate $f_{CP}$. In the case of $f_{CP} = \pi^+
\pi^-$, the time-dependent rate is given by
 \beq
 f_{\pi\pi}(\Delta
t) = \frac{e^{-|\Delta t|/ \tau_{B^0}}}{4 \tau_{B^0}} \left \{ 1+
q\cdot \left [ \spp \sin{(\Delta m_d \Delta t)} + A_{\pi\pi}
\cos{(\Delta m_d \Delta t)} \right ] \right \}  ,
 \eeq
where $\tau_{B^0}$ is the $B_d^0$ lifetime, $\Delta m_d$ is the mass
difference between the two $B_d^0$ mass eigenstates, $\Delta t
=t_{CP} -t_{tag}$ is the time difference between the tagged-$B^0$
($\bar{B}^0$) and the accompanying $\bar{B}^0$ ( $B^0$) with
opposite $b-$flavor decaying to $\pi^+ \pi^-$ at the time
$t_{CP}$, $q=+1$ ($-1$) when the tagging $B$ meson is a $B^0$
($\bar{B}^0$).
The CP-violating asymmetries $\spp$ and $A_{\pi\pi}$ have been defined as
\beq
S_{\pi\pi} = \frac{2 \im(\lpp)}{|\lpp|^2 +1 }, \ \
A_{\pi\pi} &=& \frac{|\lpp|^2 - 1}{|\lpp|^2 +1 },
\eeq
where the parameter $\lpp$ is of the form
\beq
\lpp &=& \frac{q}{p} \frac{\bar{A}}{A} = e^{2i\alpha} \frac{1 - r e^{i(\delta -\alpha)}}{
1 - r e^{i(\delta +\alpha)}},\label{eq:lpp}
\eeq
where $\delta = \delta_P-\delta_T$ is the difference of CP conserving strong
phases, while the ratio of penguin to tree amplitudes of the $B \to \pi^+ \pi^-$ decay
takes the form
\beq
r= \left |\frac{P}{T}\right | =\frac{|V_{tb}^* V_{td}|}{|V_{ub}^* V_{ud}|}
\frac{|P_{tc}|}{|T_{uc}|}, \label{eq:r}
\eeq
based on the definition of decay amplitudes in Eqs.~(\ref{eq:a1a},\ref{eq:a11}).
By explicit calculations, we find that\footnote{By a transformation of $\delta \to \pi + \delta$,
the expressions of $\spp$ and $\app$ here will become  identical with those in
Ref.\cite{lx02}.}
\beq
\spp &=& \frac{\sin{2\alpha} - 2 r \cos{\delta } \sin{\alpha}}{
1+ r^2 - 2r\cos{\delta}\cos{\alpha}}, \label{eq:sppth}\\
\app &=& \frac{- 2 r \sin{\delta} \sin{\alpha}}{
1+ r^2 - 2r\cos{\delta}\cos{\alpha}}. \label{eq:appth}
\eeq

By setting $r=0$, one would find
\beq
A_{\pi \pi}=0, S_{\pi\pi}=\sin(2\alpha),
\eeq
which clearly means that one can measure the $\sin (2\alpha)$ directly from $\bpp$ decay
in  case of neglecting the penguin contribution to this decay.
This is the reason why $\bpp$ decay was assumed to be the best channel
to measure CKM angle $\alpha$ previously. But the measurements of $B \to \pi\pi$ and $K \pi$ decays
show that the penguin contribution should be rather large. With penguin contributions, we have
$A_{\pi \pi}\neq 0$ and $\spp \neq \sin(2\alpha)$. Besides the Eq.(\ref{eq:sppth}), $\spp$
can also be defined as \cite{charles99}
\beq
\spp = \frac{1}{\sqrt{1-\app^2}}\sin(2\alpha_{eff}), \label{eq:spp-eff}
\eeq
or
\beq
\spp=\sin(2\bar{\alpha}_{eff}). \label{eq:spp-eff2}
\eeq
Here the allowed range of $\alpha_{eff}$ can be determined by the measurements of $\spp$ and
$\app$ directly. The key problem is how to determine the differences
\beq
\Delta{\alpha}= \alpha-\alpha_{eff}, \ \ \Delta{\bar{\alpha}}= \alpha-\bar{\alpha}_{eff},
\eeq
which depends on the magnitude and strong phases of the tree and  penguin amplitudes.
In this case, the CP asymmetries can not tell the size
of angle $\alpha$ directly. A method has been proposed to extract CKM angle $\alpha$
using $B^+\to \pi^+\pi^0$ and $B^0\to \pi^0\pi^0$ decays together with $\bpp$
decay  by the isospin relation \cite{gronau90}. However, it will take quite some
time for the experiments  to measure the three channels together.
In Ref.\cite{charles99}, Charles studied the ways to estimate $\Delta\alpha$
from the measured branching ratios of $B \to \pi \pi$ and $B \to K \pi$ decays.
In next section, we will extract the constraints on both $\alpha$ and $\delta$
directly by considering the latest experimental measurements about the relevant
branching ratios of $B \to \pi^+ \pi^-, \pi^+\pi^0$ and $K^0 \pi^+$ decays.

\subsection{Estimation of the ratio $r=|P_{\pi\pi}/T_{\pi\pi}|$}\label{sec-32}

From Eqs.(\ref{eq:sppth},\ref{eq:appth}), one can see that the asymmetry parameter
$\spp$ and $\app$ generally depend on only three ``free" parameters:
the CKM angle $\alpha$, the strong phase
$\delta$ and the ratio $r$. We can not solve out these
two equations with three unknown variables directly. However, we can find definite
constraint on the angle $\alpha$ and strong phase $\delta$ if the ratio $r=|P/T|$
can be determined experimentally, or at least estimated theoretically with good precision.

Up to now, many attempts have been made to estimate the ratio $r$ in various ways
\cite{gronau02,charles99,luo02,gronau90,gronau94,bbns99,bbns01}.
Three typical estimations of $r$ are the following.

In Ref.~\cite{bbns01}, Beneke et al. made a theoretical estimation of the ratio
$r$. Employing the QCD factorization approach and neglecting the annihilation amplitude,
the $B \to \pi^+ \pi^-$ decay amplitude can be written as \cite{bbns01}
\beq
{\cal A}(B^0 \to \pi^+ \pi^-) &=&
 V_{ub}^*V_{ud} \left [ a_1 +  a_4^u + a_{10}^u + r_\chi^\pi \left ( a_6^u + a_8^u \right )
 \right ] A_{\pi\pi} \non
&& +  V_{cb}^*V_{cd} \left [  a_4^c + a_{10}^c + r_\chi^\pi \left ( a_6^c + a_8^c \right )
 \right ] A_{\pi\pi} \non
&& + \left [ V_{ub}^*V_{ud} b_1 + (V_{ub}^*V_{ud} + V_{cb}^*V_{cd}) \left (
b_3 + 2b_4 -\frac{1}{2} b_3^{EW} + \frac{1}{2} b_4^{EW} \right ) \right ] B_{\pi\pi}
\label{eq:app-bbns}
\eeq
with
\beq
A_{\pi\pi} &=& - i \frac{G_F}{\sqrt{2}}(m_B^2 -m_\pi^2)F_0^{B\to
\pi}(m_\pi^2)f_\pi,\\
B_{\pi\pi} &=& - i \frac{G_F}{\sqrt{2}} f_B f_\pi^2,
\eeq
where the coefficients $a_1$ and $a_2$ describe the tree diagram contribution,
$a_i^u$ and $a_i^c$ describe the QCD penguin (for $i=4,6$) and electroweak penguin
(for $i=8,10$) contributions
induced by the penguin diagrams with internal up or charm quark  respectively,
the coefficients $b_i$ describe the weak annihilation contributions
\cite{bbns01}, and $f_B$ and $f_\pi$ are the B and $\pi$ meson decay constants,
$F_0^{B \to \pi}$ is the form factor of $B \to \pi$ transition, and the factor
$r_\chi^\pi $ describes the so-called chiral-enhancement\cite{bbns01} to $B\to \pi^+ \pi^- $
decay.  The explicit expressions of $a_i$ can be found in Ref.\cite{bbns01}.

In Ref.~\cite{bbns01}, the ``tree" and ``penguin" part of decay
amplitude (\ref{eq:app-bbns}) were defined as the amplitude proportional to
the CKM factor $ V_{ub}^*V_{ud} $ and $ V_{cb}^*V_{cd} $ respectively
(i.e. using the convention B), and the ratio took the form \cite{bbns01}
\beq
\left | \frac{P_{\pi\pi} }{T_{\pi\pi} } \right |
= - \frac{| V_{cb}^*V_{cd} |}{ |V_{ub}^*V_{ud}| }
\left | \frac{  a_4^c + a_{10}^c + r_\chi^\pi \left ( a_6^c + a_8^c \right )
+ r_A \left [b_3 + 2b_4 -\frac{1}{2} \left ( b_3^{EW} -b_4^{EW} \right ) \right ]      }{
a_1 + a_4^u + a_{10}^u + r_\chi^\pi \left ( a_6^u + a_8^u \right )
+ r_A \left [b_1 + b_3 + 2b_4 -\frac{1}{2} \left ( b_3^{EW} -b_4^{EW} \right ) \right ] } \right |
\label{eq:r-bbns}
\eeq
where $r_A=B_{\pi\pi}/A_{\pi\pi} \approx 0.003$. The corresponding numerical result
was  \cite{bbns01}
\beq
r= \left | \frac{P_{\pi\pi} }{T_{\pi\pi} } \right | =  0.285 \pm 0.076\ \ (0.259 \pm 0.068),
\label{eq:r2-bbns}
\eeq
when the weak annihilation contributions are included or not.

The magnitude of the ratio $r$ can also be estimated by considering the
measured branching ratios for the tree dominated $B\to \pi^+ \pi^0$ decay
and the pure penguin channel $B \to K^0 \pi^+$. But because of the unitarity
relation of the CKM matrix, one can define the tree and penguin  amplitude
in different ways, i.e. choose different convention in one's calculation.
For $B \to \pi^+ \pi^-$ decay, three are three different conventions, as defined
in Eqs.(\ref{eq:a1a}-\ref{eq:a1c}). For $B \to \pi^+ \pi^0$ and $K^0 \pi^+$ decay
modes, we meet  the similar situation.

The first convention considered here is the Luo and Rosner (LR) convention.
In Ref.\cite{luo02}, Luo and Rosner defined the strangeness-preserving
$\bar{b} \to \bar{d}$ penguin amplitude $ P_{\pi\pi}$ in the $B \to \pi^+ \pi^-$ decay
as the amplitudes proportional to the CKM combination $V_{td}V_{tb}^*$ (i.e. the convention A
as given in Eq.(\ref{eq:a1a})), while set the $\bar{b} \to \bar{s}$ penguin amplitude
$ P_{K\pi}$ in the pure penguin $B^+ \to K^0 \pi^+ $
decay as the amplitudes proportional to $V_{ts}V_{tb}^*$. By quoting all decay rates in
units of ($B$ branching ratio $\times 10^{6}$) and using the world average available at
that time \cite{luo02}, they found that
\beq
\calb (B^+ \to K^0 \pi^+) = \frac{\tau^0}{\tau^+} |P_{K\pi}|^2  = 17.2 \pm 2.4,
\eeq
which leads to
\beq
|P_{K\pi}| =4.02 \pm 0.28,
\eeq
and
\beq
|P_{\pi\pi}| \simeq \left |\frac{V_{td}}{V_{ts}} \right | |P_{K\pi}| = 0.71 \pm 0.14.
\eeq

By using the theoretical estimation of the form factor shapes and the measurement
of the spectrum of $B \to \pi l \nu$ near $q^2=0$ where $q^2$ is the squared effective
mass of the $l\nu$ system, Luo and Rosner \cite{luo02} estimated  the ``tree" part of
the $B \to \pi^+ \pi^-$ decay rate  and then obtained the value of the parameter $T_{\pi\pi}$
\beq
|T_{\pi\pi}| = 2.7 \pm 0.6,
\eeq
which leads to
\beq
r= \left | \frac{P_{\pi\pi}}{T_{\pi\pi}}\right | = 0.26 \pm 0.08. \label{eq:r-lr}
\eeq

The second convention considered here is the Gronau and Rosner (GR) convention.
In Ref.\cite{gronau02}, Gronau and Rosner defined the penguin amplitude
$ P_{\pi\pi}$ of $B \to \pi^+ \pi^-$ decay as the
amplitude proportional to the CKM combination $V_{cd}V_{cb}^*$ (i.e. the convention B
as given in Eq.(\ref{eq:a1b})), and set the $\bar{b} \to \bar{s}$ penguin amplitude
$ P_{K\pi}$ in the pure penguin $B^+ \to K^0 \pi^+ $ decay as the amplitude proportional
to $V_{cs}V_{cb}^*$, they then found numerically that
\beq
|P_{\pi\pi}| = \frac{|V_{cd}|}{|V_{cs}|} \frac{f_\pi}{f_K} |P_{K\pi}|
= \lambda \frac{f_\pi}{f_K} |P_{K\pi}| \simeq
0.74 \pm 0.05, \label{eq:pp-gr}
\eeq
where $f_\pi / f_K\approx 0.84 $ is the $SU(3)$ breaking factor.
Combining this result with
\beq
|T_{\pi\pi}| =2.7 \pm 0.6, \label{eq:t-gr}
\eeq
estimated from the measured branching ratio $\calb (B^+ \to  \pi^+ \pi^0)
= (4.6 \pm 2.0 )\times 10^{-6} $ and the assumed $Re(C_{\pi\pi}/T_{\pi\pi}) =0.1$ \cite{rosner01},
they  found the value of the ratio $r$ \cite{gronau02}
\beq
r= \left | \frac{P_{\pi\pi}}{T_{\pi\pi}}\right | = 0.276 \pm 0.064.
\label{eq:r-gr}
\eeq

From the values as given in Eqs.(\ref{eq:r2-bbns},\ref{eq:r-lr},\ref{eq:r-gr}),
one can see that the ratio $r$ obtained by three different estimations are
in good agreement within $1\sigma$ error. From the analysis in this section and the
estimations in next section where the newest data will be considered, we believe that
it is reasonable to set $r= 0.3 \pm 0.1 $ in our calculations. In next section,
we find that the measurements of
relevant branching ratios prefer a ratio $r$ larger than $0.16$. In Belle's paper,
they assumed that $r=0.3 \pm 0.15$ to draw the constraint on the CKM angle
$\alpha$ from their latest measurements of $\spp$ and $\app$ \cite{0301032}.

\subsection{Constraints on $\alpha$ and $\beta$  }

Now we are ready to extract out $\alpha$ by comparing the theoretical prediction
of $\spp$ and $\app$ with the measured results. As discussed previously \cite{gr02},
there may exist some discrete ambiguities between $\delta$ and $\pi - \delta $ for the
mapping of $\spp$ and $\app$ onto the $\delta-\alpha$ plane. We will consider
the effects of such discrete ambiguity.

At first, the positiveness of the measured $\app$ at $2\sigma$ level and the fact that
$\sin{\alpha} >0$ for $0 < \alpha < 180^\circ$ implies that
$\sin{\delta}$ should be negative, the range of $0  \leq  \delta  \leq 180^\circ $
is therefore excluded under the parametrization of decay amplitude of $B \to \pi^+ \pi^-$
as given in Eqs.(\ref{eq:a11},\ref{eq:a21}). Consequently, only the range of
$-180^\circ <  \delta < 0^\circ$ need to be considered here.

For the special case of $\delta=-90^\circ$, the discrete ambiguity between $\delta$ and
$\pi-\delta$ disappears and the expressions of $\spp$ and $\app$ can be
rewritten as
\beq
\spp &=& \frac{ \sin{2\alpha} }{ 1 + r^2 },\label{eq:spp90} \\
\app &=& \frac{  2 r \sin{\alpha}}{ 1 + r^2 }, \label{eq:app90}
\eeq
which leads to the common allowed range of angle $\alpha$
\beq
98^\circ \leq  \alpha \leq 118^\circ
\eeq
for $\spp=-0.49\pm 0.27$,  $\app=0.51\pm 0.19$ and $r=0.30 \pm 0.10$.

\begin{figure}[hbt]
\vspace{-40pt}
\begin{minipage}[]{\textwidth}
\centerline{\epsfxsize=\textwidth \epsffile{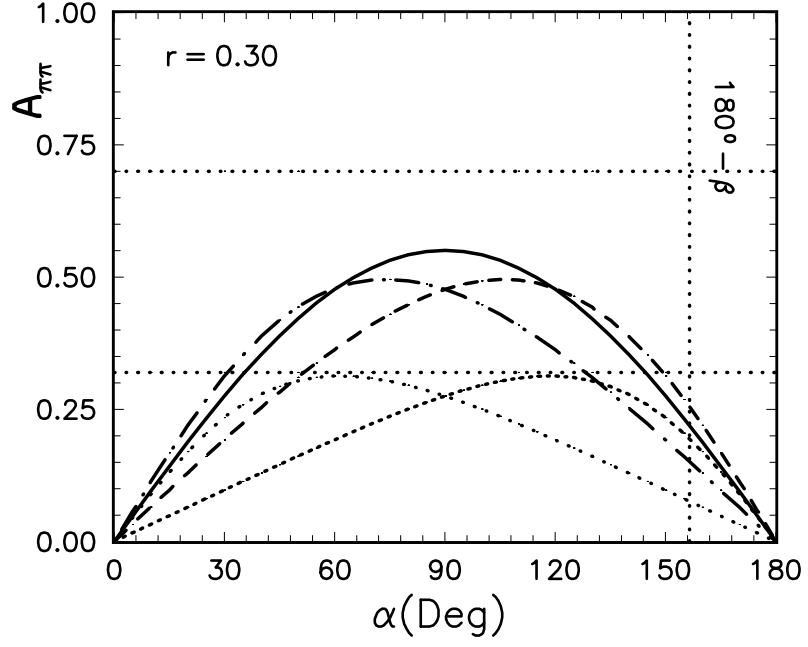}}
\vspace{-20pt}
\caption{Plots of $\app$ vs angle $\alpha$ for $r=0.30$.
The five curves correspond to $\delta =
-30^\circ$ (dots curve), $-60^\circ$ (dot-dashed curve), $-90^\circ$ (solid curve),
$-120^\circ$ (short-dashed curve) and $-150^\circ$ (tiny-dashed curve), respectively.
The band between two horizontal dots lines shows the allowed range
$ 0.32 \leq \app^{exp} \leq 0.70$ at $1\sigma$ level.
The vertical dots line refers to the physical limit $\alpha \leq 156.4^\circ$. }
\label{fig:fig3}
\end{minipage}
\end{figure}

For an arbitrary value of $\delta$ in the region of $ - 180^\circ < \delta < 0^\circ$, the
constraint on the angle $\alpha$ will change.
In Fig.~\ref{fig:fig3}, we show the $\alpha$ dependence of $\app$ for $r=0.3$
and  $\delta = -30^\circ$ (dots curve), $-60^\circ$ (dot-dashed curve), $-90^\circ$ (solid curve),
$-120^\circ$ (short-dashed curve) and $-150^\circ$ (tiny-dashed curve), respectively.
The band between two horizontal dots lines shows the allowed range
$ 0.32 \leq \app^{exp} \leq 0.70$. The vertical dots line refers to the physical
limit $\alpha \leq 180^\circ -\beta$ with $\beta=23.6^\circ$. It is easy to see
from Fig.~\ref{fig:fig3} that the regions allowed by the measured $\app$ alone at $1\sigma$ level
are
\beq
 30^\circ \leq \alpha \leq 150^\circ, \\
-150^\circ \leq \delta \leq -30 ^\circ,
\eeq
for $r=0.30$ and $\app =0.51 \pm 0.19$.

\begin{figure}[]
\vspace{-40pt}
\begin{minipage}[]{\textwidth}
\centerline{\epsfxsize=\textwidth \epsffile{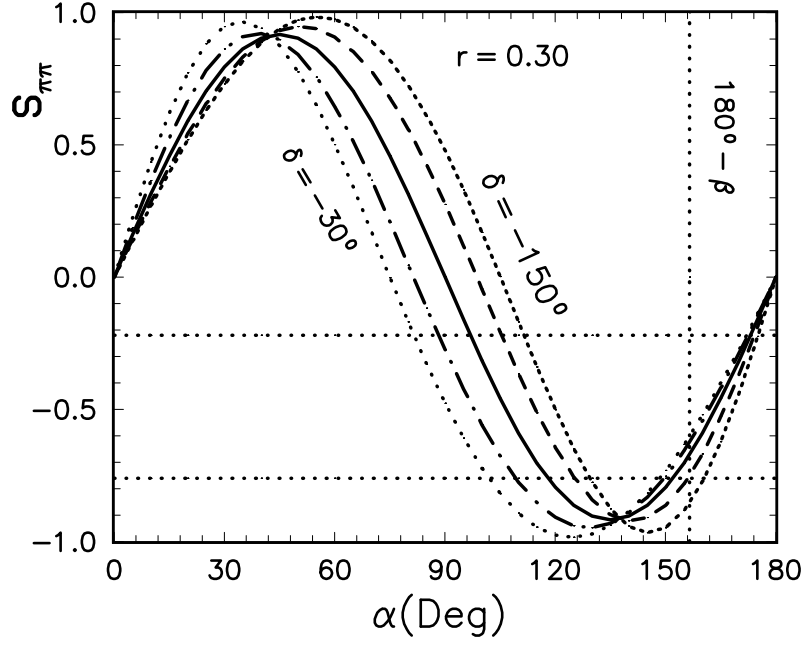}}
\vspace{-20pt}
\caption{Plots of $\spp$ vs angle $\alpha$ for $r=0.30$ and $-150^\circ \leq \delta
\leq -30^\circ$. From the left to the right the five curves correspond to $\delta =
-30^\circ$, $-60^\circ$, $-90^\circ$, $-120^\circ$ and $-150^\circ$, respectively.
The band between two horizontal dots lines shows the allowed range
$ -0.76 \leq \spp^{exp} \leq -0.22$ at $1\sigma$ level.
The vertical dots line refers to the physical limit $\alpha \leq 156.4^\circ$. }
\label{fig:fig4}
\end{minipage}
\end{figure}

In Fig.\ref{fig:fig4}, we show the $\alpha$ dependence of $\spp$ for $r=0.3$
and  $\delta = -30^\circ$ (dots curve), $-60^\circ$ (dot-dashed curve), $-90^\circ$ (solid curve),
$-120^\circ$ (short-dashed curve) and $-150^\circ$ (tiny-dashed curve), respectively.
The band between two horizontal dots lines shows the measured value of $ \spp =-0.49 \pm 0.27$.
The vertical dots line shows the physical limit $\alpha \leq 180^\circ -\beta$ with
$\beta=23.6^\circ$. From this figure, we find that
the range of $82^\circ \leq \alpha \leq 129^\circ$ is allowed by the
measured $\spp$ at $1\sigma$ level.

\begin{figure}[t]%fig.5
\vspace{-100pt}
\begin{minipage}[]{\textwidth}
\centerline{\epsfxsize=\textwidth \epsffile{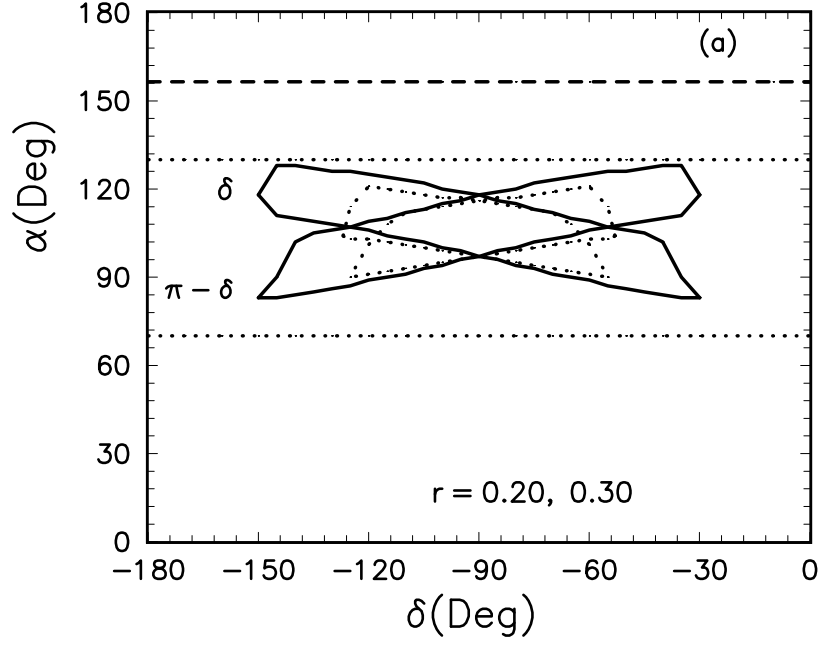}}
\vspace{-50pt}
\centerline{\epsfxsize=\textwidth \epsffile{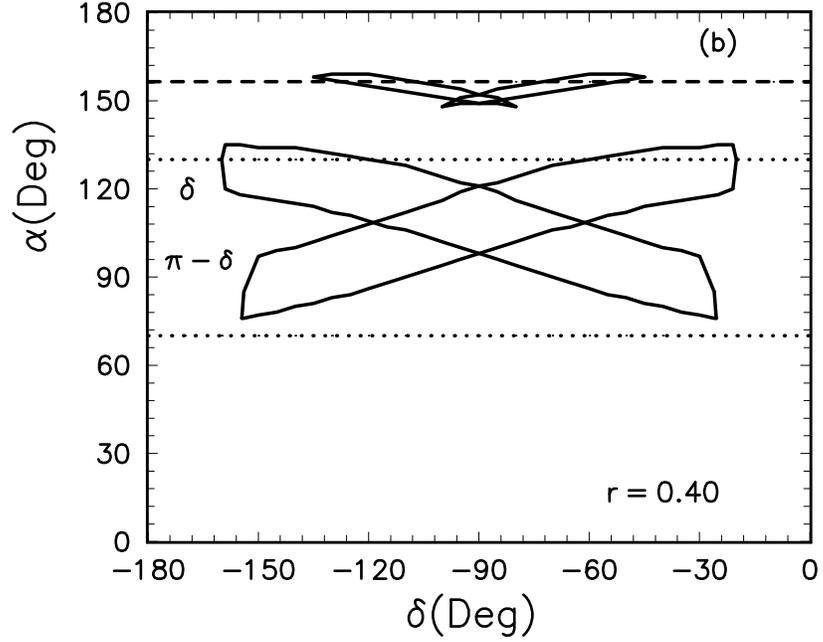}}
\vspace{-30pt}
\caption{Contour plots of the asymmetries $\spp$ and $\app$ versus the CKM angle
$\alpha$ and the strong phase $\delta$
for $r=0.2$ (the dots circles in (a)), $0.3$ (the solid circles in (a)), and $0.4$(b),
respectively. The upper short-dashed line
shows the physical upper limit, while the band between two horizontal dots lines
shows the global fit result: $70^\circ \leq \alpha \leq 130^\circ$.}
\label{fig:fig5}
\end{minipage}
\end{figure}

Fig.~\ref{fig:fig5} shows the contour plots of the CP asymmetries $\spp$ and
$\app$ versus the strong phase $\delta$ and CKM angle $\alpha$ for $r=0.20$
(the dashed circles in (a)), $0.30$ (the solid circles in (a)) and $0.40$
(b), respectively. The regions inside each circle
are allowed by both $\spp^{exp} = -0.49 \pm 0.27$ and $\app^{exp} = 0.51 \pm 0.19$
for given $r$. The discrete ambiguity between
$\delta$ and $\pi -\delta$ are also shown in Fig.\ref{fig:fig5}.
For $\delta =-90^\circ$, such discrete ambiguity disappear. The horizontal short-dashed
line is the physical limit $\alpha \leq 156.4^\circ$, while the band between two
horizontal dots lines shows the allowed region of $30^\circ \leq \alpha \leq 130^\circ$
from the global fit.

The constraint on the CKM angle $\alpha$ and the strong phase $\delta$
can be read off directly from Fig.\ref{fig:fig5}.
Numerically, the allowed regions for the CKM angle $\alpha$ and
the strong phase $\delta$ are
\beq
90^\circ \leq \alpha \leq 121^\circ, \ \  -127^\circ \leq \delta \leq -55^\circ,
\label{eq:z20}
\eeq
for $r=0.2$,  and
\beq
83^\circ \leq \alpha \leq 128^\circ,  \ \  -150^\circ \leq \delta \leq -33^\circ,
\label{eq:z30}
\eeq
for $r =0.3$,  and finally
\beq
&& 76^\circ \leq \alpha \leq 135^\circ \bigvee -160^\circ \leq \delta \leq -26^\circ,
\label{eq:z40a} \\
&& 148^\circ \leq \alpha \leq 158^\circ \bigvee -135^\circ \leq \delta \leq -45^\circ,
\label{eq:z40b}
\eeq
for $r =0.4$. There is a twofold ambiguity for the determination of angle $\alpha$ for
$r= 0.4$, but the second region of $\alpha$ in Eq.~(\ref{eq:z40b}) is  strongly
disfavored by  the global fit result as illustrated in Fig.\ref{fig:fig5}
and will be dropped.

\begin{figure}[]
\vspace{-40pt}
\begin{minipage}[]{\textwidth}
\centerline{\epsfxsize=\textwidth \epsffile{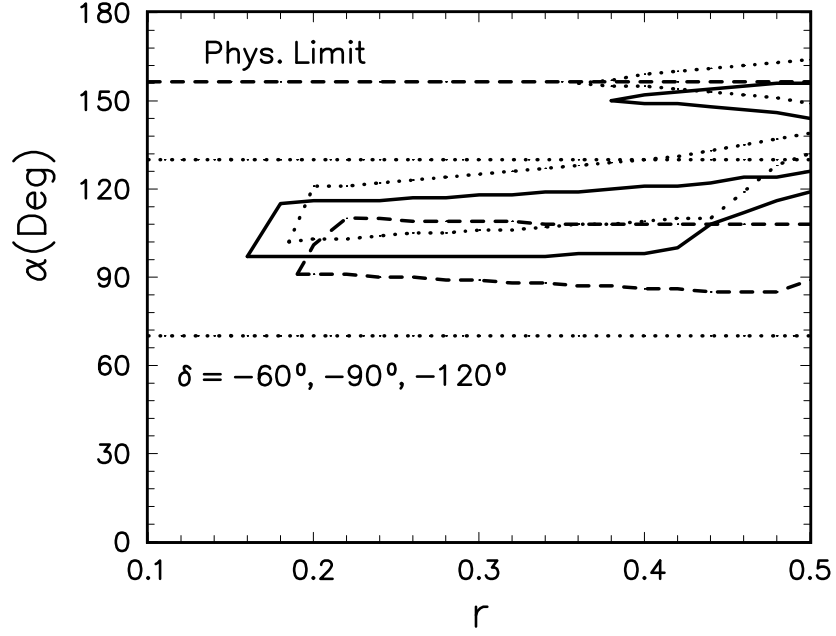}}
\vspace{-20pt}
\caption{Contour plots of the asymmetries $\spp$ and $\app$ versus the CKM angle $\alpha$
and the ratio $r=|P/T|$ for $\delta=-60^\circ$ (dots curves), $-90^\circ$ (solid curves)
and $-120^\circ$ ( short-dashed curves), respectively. The regions inside the semi-closed
curves are allowed by the data. The upper short-dashed line shows the physical limit
$\alpha \leq 156.4^\circ$,  while the band between two horizontal dots lines
shows the global fit result: $70^\circ \leq \alpha \leq 130^\circ$.}
\label{fig:fig6}
\end{minipage}
\end{figure}

Obviously, the constraints on $\alpha$ strongly depend on the value of $r$. In
Fig.\ref{fig:fig6} we draw the contour plots of the CP asymmetries $\spp$ and
$\app$ in the $r-\alpha$ plane for $\delta=-60^\circ$
(short-dashed curves), $-90^\circ$ ( solid curves) and $-120^\circ$ (dots curves),
respectively. The regions inside each semi-circle
are allowed by both $\spp^{exp} = -0.49 \pm 0.27$ and $\app^{exp} = 0.51 \pm 0.19$
(experimental $1\sigma$ allowed ranges when the FDG scaling factor is not considered).
The upper short-dashed line show the physical
limit $\alpha \leq 156.4^\circ$, the band between two horizontal dots lines
shows the global fit result.
From this figure and the numerical calculations we find a lower limit on $r$,
\beq
r \geq 0.16 \label{eq:limit-r}
\eeq
for the whole possible ranges of $\delta $ and $\alpha$, which agrees well with previous
estimations.

In Ref.\cite{0301032}, based on their own newest measurement, Belle Collaboration
presented  the allowed region of $\alpha$ ( at $95.5\%$ C.L.)
\beq
78^\circ \leq \alpha \leq 152^\circ \label{eq:belle-l}
\eeq
for $\beta =23.5^\circ$ and $0.15 \leq r \leq 0.45$.

In our analysis, we used the weighted average of BaBar and Belle's measurements
of CP-violating asymmetries and $r=0.3 \pm 0.1$ as the experimental and theoretical input.
Our result is in good agreement with Belle's constraint (\ref{eq:belle-l}) where only
their own measurement of $\spp$ and $\app$ were used.

One can see from above analysis that if we take the average $\spp=-0.49 \pm 0.27$ and
$\app=0.51 \pm 0.19 $ as the experimental input, and use the
theoretically fixed ratio $ r = 0.3 \pm 0.1$, we find a strong constraint on $\alpha$,
\beq
76^\circ \leq \alpha \leq 135^\circ. \label{eq:al-a}
\eeq
This range of $\alpha$  is  well consistent with the global fit
result\cite{hocker01,he01,ckmfit,pdg2002}. On the other hand, the constraint on the strong
phases $\delta $ from the measured $\spp$ and $\app$ is still rather weak.

As discussed in the introduction, the measurements of $\spp$ and $\app$ as reported by
BaBar and Belle Collaboration are not fully consistent. We should consider the case of
using  the weighted-average with the errors magnified by the PDG scaling factor,
as already given in Eq.(\ref{eq:spp-exp})
\beq
\spp &=& -0.49 \pm 0.61, \app = +0.51 \pm 0.23. \label{eq:spp-exp2}
\eeq

\begin{figure}[t]%fig.5
\vspace{-100pt}
\begin{minipage}[]{\textwidth}
\centerline{\epsfxsize=\textwidth \epsffile{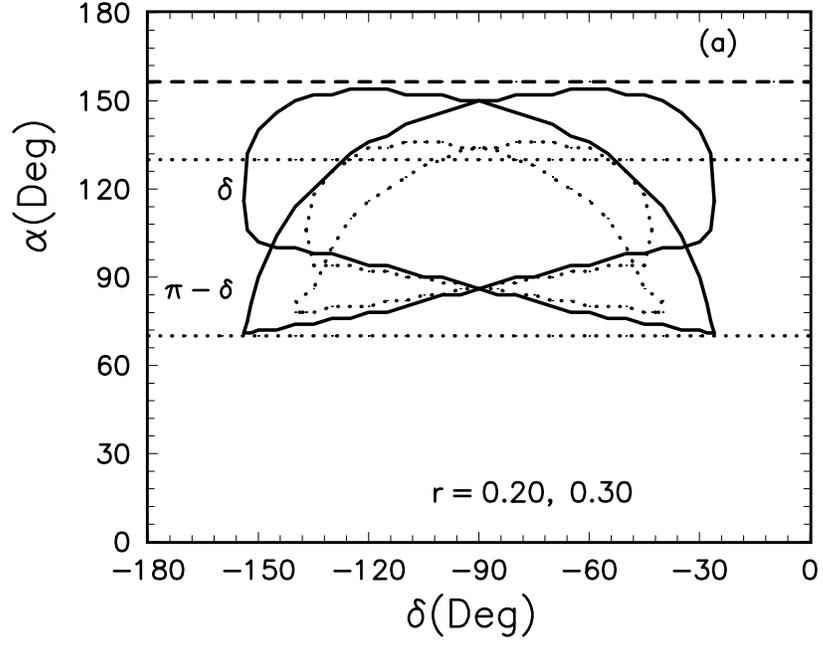}}
\vspace{-50pt}
\centerline{\epsfxsize=\textwidth \epsffile{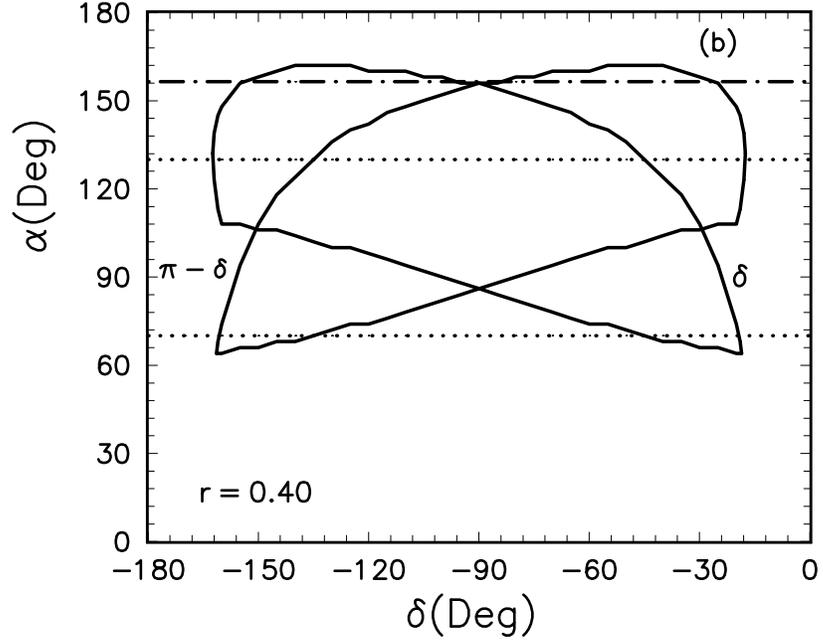}}
\vspace{-30pt}
\caption{The same as Fig.\ref{fig:fig5}, but using $\spp=0.49 \pm 0.61$ and
$\app=0.51 \pm 0.23$ as the experimental input. }
\label{fig:fig7}
\end{minipage}
\end{figure}

In Fig.\ref{fig:fig7}, we draw the contour plots of the CP asymmetries $\spp$
and $\app$  versus the strong phase $\delta$ and the CKM angle $\alpha$ by
using $\spp=-0.49\pm 0.61$ and $\app=0.51 \pm 0.23$. The allowed regions are
then
\beq
78^\circ \leq \alpha \leq 136^\circ, \ \  -137^\circ \leq \delta \leq -43^\circ
\label{eq:z20c}
\eeq
for $r=0.2$,  and
\beq
71^\circ \leq \alpha \leq 154^\circ,  \ \  -154^\circ \leq \delta \leq -26^\circ
\label{eq:z30c}
\eeq
for $r =0.3$,  and finally
\beq
64^\circ \leq \alpha \leq 162^\circ \bigvee -162^\circ \leq \delta \leq -18^\circ
\label{eq:z40c}
\eeq
for $r =0.4$. It is easy to see from figures \ref{fig:fig5} and \ref{fig:fig7} that  the
constraints on $\alpha$ become less restrictive now.

In next section, we try to draw new constraints on the angle $\alpha$ and $\delta $ from the
measured CP-averaged branching ratios of $B \to \pi^+ \pi^-, \pi^+ \pi^0$ and
$B \to K^0\pi^+$ decays.

\section{Information from measured branching ratios}\label{sec-4}

As listed in \tab{br}, the CP-averaged branching ratios of $B \to \pi^+ \pi^-, \pi^+ \pi^0$
and $B \to K^0 \pi^+$ decay modes have been measured with rather good precision.
From these measured ratios one can infer that (a) the ratio of tree to penguin amplitudes
$r=|P_{\pi\pi}/T_{\pi\pi}|$; and
(b) the constraint on the angle $\alpha$ and strong phase $\delta$. Of course, the
information from measured $\calb (B \to K \pi)$ and $\calb(B \to \pi \pi)$ can
be used as a crosscheck for the results obtained in the last section.

\subsection{The measured branching ratios and the ratio $r$ }

In this paper, we use the following assumptions in addition to the SM \cite{charles99}:
\begin{enumerate}
\item
$SU(2)$ isospin asymmetry of the strong interactions.

\item
The breaking of $SU(3)$ flavor symmetry of the strong interactions is described
by the ratio of $f_\pi/f_K$ with $f_\pi =133 MeV$ and $f_K = 158 MeV$ as given in
\tab{input}.

\item
Neglect of the suppressed annihilation diagrams and the electroweak
penguin contributions.

\item
Neglect of the penguin contributions which is proportional to
$V_{us}V_{ub}^*$  in the $B^+ \to K^0 \pi^+$ decay amplitude.

\end{enumerate}

Following Ref.\cite{gronau94}, the flavor-$SU(3)$ decomposition of
$B \to \pi^+ \pi^-, \pi^+ \pi^0$ and $K^0 \pi^+ $ decay amplitudes can be in general
written as
\beq
A(B^+ \to \pi^+ \pi^0)&=& -\frac{1}{\sqrt{2}} \left (  T + C +P_{EW} + P_{EW}^{C}\right ),
\label{eq:pppz}\\
A(B^0 \to \pi^+ \pi^-)&=& -\left (  T + P  -\frac{2}{3}P_{EW}^{C}\right ),
\label{eq:pppm}\\
A(B^+ \to K^0 \pi^+) &=&  P^{'} -\frac{1}{3}P_{EW}^{'C}\, , \label{eq:k0pp}
\label{eq:kppm}
\eeq
where the unprimed amplitudes denote strangeness-preserving ($\Delta S=0$) $ b \to d$ decays,
the primed amplitudes denote strangeness changing ($|\Delta s |=1$) $b \to s $ decays.
The amplitude $T$ is  color-favored tree amplitude, $C$ is the color-suppressed
tree amplitude, $P$ and $P^{'}$ are gluon penguin amplitudes, and $P_{EW}$
and $P_{EW}^{'C}$ are the color-allowed and color-suppressed electroweak penguins (EWP)
\footnote{For a detailed discussion
about the flavor $SU(3)$ decomposition of two-body hadronic B meson decays, one can see
Refs.\cite{gronau94,rosner01} and references therein.}.
From previous studies\cite{gronau94,rosner01,he95}, we get to know that
\begin{itemize}
\item
The electroweak penguin component in $B^+ \to \pi^+ \pi^0$ decay (which is purely
$I=2$) should be between $\calo (\lambda^2)$ and $\calo (\lambda^3)$ of the dominant
$T$ contribution\cite{gronau94}, which is in agreement with Deshpande and He's
estimation \cite{he95}: $|P_{EW}/T| \approx 1.6\% |V_{td}/V_{ub}|$.
For the contribution due to $P_{EW}^{'C}$, it is in fact
much smaller than that of $P_{EW}$ because of further color-suppression.
The EWP contributions to $B^+ \to \pi^+ \pi^0, \pi^+ \pi^-$ and $K^0 \pi^+$ decays are
therefore very small and can be neglected safely.

\item
The effects of EWP amplitudes on the extraction of the CKM angle $\alpha$ are
at most of order $\lambda^2$ and are negligible\cite{gronau94}.

\end{itemize}

By employing the assumptions 1-4, one can find the constraint on the size of $r$
from the measured CP-averaged branching ratios as listed in \tab{br}. This is
shown below.

\subsection{Current estimations of $|T|$ and $|P|$}

As discussed in last section, the ratio $r=|P/T|$ can be estimated experimentally from the
measured branching ratios of the tree dominated $B \to \pi^+ \pi^0 $ decay and the pure
penguin process $B \to K^0 \pi^+$. In this subsection, we update the estimation of the ratio $r$
by considering the latest measurements of the two branching ratios.

When the electroweak penguin contribution is neglected, the $B^+ \to \pi^+ \pi^0$ decay depends
on tree amplitude $ T + C $ only. From  the well measured branching ratio as given in \tab{br},
one can determine the magnitude of $|T + C|$. If we quote all decay rates
in units of ($B$ branching ratio $\times 10^{6}$) \cite{gronau02,luo02},
the measured $B \to \pi^+ \pi^0 $ branching ratio as given in \tab{br} implies
\beq
\calbb (B^+ \to \pi^+ \pi^0) &=&\frac{1}{2}\left [ \calb(B^+ \to \pi^+ \pi^0)
+ \calb(B^- \to \pi^- \pi^0) \right ]\non
&=& \frac{1}{2} \frac{\tau^+}{\tau^0} |T + C|^2 = 5.3 \pm 0.8.  \label{eq:t1}
\eeq
We have to estimate the relative strength of $T$ and $C$ contribution before
we can fix  $|T|$ from the measured branching ratio. In Ref.\cite{rosner01},
the author assumed $Re[C/T]=0.1$. We here use the QCD factorization approach to
estimate the value of $|C/T|$.

Under the QCD factorization approach, the decay amplitudes of $B^+ \to \pi^+
\pi^0$ and $B^+ \to K^0 \pi^+$ can be written as \cite{bbns01}
\beq
A(B^+ \to \pi^+ \pi^0 ) & =&  - i \frac{G_F}{\sqrt{2}}(m_B^2 -m_\pi^2)F_0^{B\to
\pi}(m_\pi^2)f_\pi \cdot V_{ub}^* V_{ud} \left [ a_1 + a_2 \right ],
\label{eq:pmp0}
\eeq
where $a_1$ and $a_2$ describe the tree the color-suppressed tree contributions,
the small electroweak penguin has been neglected.
Using the coefficients $a_{1,2}$ at the NLO level as given in
Ref.\cite{bbns01}, we find numerically that
\beq
|C/T| \approx |a_2/a_1| \approx 0.10
\eeq
for $ m_b/2 \leq \mu \leq 2m_b$ with $m_b=4.6$ GeV. It is therefore reasonable to set
\beq
|C/T|=0.10 ^{+0.10}_{-0.05} \label{eq:rct}
\eeq
in our analysis. From Eqs.(\ref{eq:t1}) and (\ref{eq:rct}), we then find that
\beq
|T_{\pi^+\pi^0}|= 2.8 \pm 0.2 (\Delta Br) \pm 0.2 ( \Delta |C/T| )
= 2.8  \pm 0.3  \label{eq:tf}
\eeq
for $\calb (B \to \pi^+ \pi^0)=(5.3 \pm 0.8)\times 10^{-6} $ and
$\tau^+/\tau^0=1.083\pm 0.017$ \cite{pdg2002}.
This estimation of $T_{\pi^+\pi^0}$ agrees well with previous results
$|T|=2.7 \pm 0.6$ \cite{luo02,rosner01} but with smaller error due to the improvement
of the data. From the assumption of the isospin symmetry of strong interaction, we have
\beq
|T_{\pi^+\pi^-}|=|T_{\pi^+\pi^0}|.
\eeq
For the sake of simplicity, we use terms $|T|$ and $|P|$ to denote $|T_{\pi^+\pi^-}|$ and
$|P_{\pi^+\pi^-}|$ without further specification.

As discussed in last section, the strangeness-changing QCD penguin amplitude $P'=P_{K\pi}$
can be determined from the measured branching ratio $\calbb (B^+ \to K^0 \pi^+)$ as
given in \tab{br}. In units of $B$ branching ratio $\times 10^{6}$, we find
\beq
\calbb (B^+ \to K^0 \pi^+) = \frac{\tau^+}{\tau^0}|P_{K\pi}|^2 =  19.7 \pm 1.5,
\label{eq:pp1}
\eeq
which leads to
\beq
|P_{K\pi}| =  4.3 \pm 0.2 .\label{eq:pp2}
\eeq

Using Gronau and Rosner convention \cite{gronau02} to define $P_{\pi\pi}$,
we find numerically that
\beq
|P| &=& \lambda  \frac{f_\pi}{f_K} |P_{K\pi}| = 0.80 \pm 0.04 \label{eq:pf1}
\eeq
for $f_\pi=0.133$ GeV and $f_K=0.158$ GeV, which leads to the value
\beq
r = |P/T| = 0.29 \pm 0.04\label{eq:ratio4}
\eeq
for $T=2.8 \pm 0.3 $ as given in Eq.(\ref{eq:tf}).

Using the  Luo and Rosner convention \cite{luo02} to define $P_{\pi\pi}$,
we find numerically that
\beq
|P| &=& \lambda |1-\rhob -i\etab| |P_{K\pi}| = 0.81\pm 0.19 \label{eq:pf2}
\eeq
for $\lambda=0.2196\pm 0.0026$, $\rhob =0.22\pm 0.10$ and $\etab=0.35 \pm 0.05$ as
given in Ref.\cite{pdg2002}. The ratio $r$ therefore takes the value of
\beq
r = |P/T| = 0.29 \pm 0.07 \label{eq:ratio5}
\eeq
for $T=2.8 \pm 0.3 $ as given in Eq.(\ref{eq:tf}).

The ratios $r$ in Eqs.(\ref{eq:ratio4}) and (\ref{eq:ratio5})  agree very well with
those given in previous papers \cite{gronau02,luo02,bbns01}, but with smaller errors
because of the improvement of the data.
We believe that it is reasonable and conservative to set $r=0.3\pm 0.1$ in our
calculations.

\subsection{Constraint on $\alpha$ and $\delta$ from measured branching ratios}

Using the expressions of the decay amplitudes as given in
Eqs.(\ref{eq:a11},\ref{eq:a21}), the CP-averaged branching
ratio of $B \to \pi^+ \pi^-$ decay can be written as
\beq
\calbb (B^0 \to \pi^+ \pi^-)=
|T|^2 + |P|^2 - 2|T||P| \cos{\delta} \cos{\alpha}.\label{eq:br-pp1}
\eeq
The world average of experimental measurements ( in units of $B$ branching ratio $\times 10^{6}$ )
implies
\beq
\calbb (B^0 \to \pi^+ \pi^-)=
|T|^2 + |P|^2 - 2|T||P| \cos{\delta} \cos{\alpha} = 4.6 \pm 0.4.\label{eq:br-pppm}
\eeq
It is easy to find that
\beq
|T|^2 + |P|^2 = 8.5 \pm 1.3  > 4.6\pm 0.4 \label{eq:tps}
\eeq
for $|T| = 2.8 \pm 0.3 $ and $|P| = 0.81 \pm 0.19$.
In other words, the tree and penguin amplitudes in
$B^0 \to \pi^+ \pi^-$ decay are indeed interfering destructively at
$2.6\sigma$ level. Consequently, $\cos\alpha$ and $\cos\delta$ in Eq.(\ref{eq:br-pppm})
should have the same sign. By scanning the whole ranges of $|T| = 2.8 \pm 0.3 $,
$|P| = 0.81 \pm 0.19$ and $\calb (B^0 \to \pi^+ \pi^-)= 4.6\pm 0.4$, we find the lower limit
\beq
\cos{\alpha} \cos{\delta} &=& \frac{|T|^2 + |P|^2 -(4.6\pm 0.4)}{2|T||P|} \geq 0.45,
\label{eq:cc}
\eeq
and more specifically,
\beq
&& |\cos{\alpha}| \geq 0.45, \label{eq:cca}\\
&& |\cos{\delta}| \geq 0.45. \label{eq:ccd}
\eeq
The minimum value of $\cos{\alpha} \cos{\delta}=0.45 $ corresponds to the choice of
$|T| = 2.5, |P| = 1.0$ (i.e. $r=|P/T|=0.4$) and $\calbb (B^0 \to \pi^+ \pi^-)= 5.0$.

Within the physical ranges of $0^\circ < \alpha < 180^\circ $ and
$-180^\circ \leq  \delta \leq 0^\circ $, the inequality (\ref{eq:cc})
leads to two sets of solutions,
\beq
{\rm I: }&& \ \ -63^\circ \leq \delta  \leq 0^\circ, \; \ \ \; 0^\circ \leq \alpha \leq 63^\circ,
\label{eq:dal-11} \\
{\rm II: }&& \ \ -180^\circ \leq \delta  \leq -117^\circ, \; \ \ \; 117^\circ \leq \alpha \leq 180^\circ.
\label{eq:dal-12}
\eeq
It is easy to see that the regions
\beq
 63^\circ \leq  \alpha  \leq  117^\circ, \ \
 -117^\circ \leq  \delta  \leq  -63^\circ,
\eeq
are excluded, which  is a new and important information
obtained from the measurements of the branching ratios of $B\to \pi^+ \pi^-, \pi^+ \pi^0$
and $B \to K^0 \pi^+$ decays.
Further improvement of the data will help us to narrow the allowed ranges for
both $\alpha$ and $\delta$. Considering the direct measurement of the angle $\beta$
and the bounds on the angle $\gamma$ from global fit, the case of $\alpha < 63^\circ$ is
strongly disfavored, while the range of
\beq
117^\circ \leq  \alpha  \leq  156.4^\circ \label{eq:limit-tpb}
\eeq
is still allowed by direct measurement of the angle $\beta$ and the data of relevant
branching ratios.

\begin{figure}[t]
\vspace{-40pt}
\begin{minipage}[]{\textwidth}
\centerline{\epsfxsize=\textwidth \epsffile{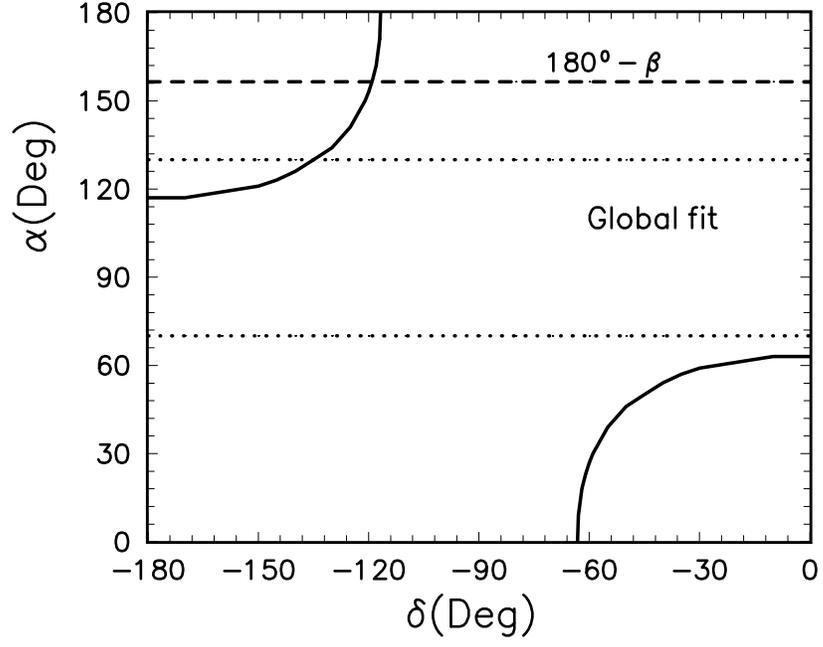}}
\vspace{-20pt}
\caption{Contour plots of the branching ratio $\calbb (B \to \pi^+ \pi^-)$ versus the
CKM angle $\alpha$ and the strong phase $\delta$. The regions at the upper-left
and the lower-right corner bounded by solid curves are still allowed by the inequality
$\cos{\alpha} \cos{\delta} \geq 0.45$. }
\label{fig:fig8}
\end{minipage}
\end{figure}

In Fig.\ref{fig:fig8} we draw the contour plots of Eq.~(\ref{eq:cc}) in the $\delta-\alpha$
plane. The two regions at the upper-left and lower-right corner bounded by solid curves
are still allowed by the constraint $\cos{\alpha} \cos{\delta} \geq 0.45$.
One can see from Fig.~\ref{fig:fig8} that  the first
solution (i.e. the lower-right corner region ) as specified in Eq.(\ref{eq:dal-11}) is
strongly disfavored by the global fit result, $70^\circ \lesssim \alpha \lesssim
130^\circ$, as illustrated by the band between to dots lines
in figure \ref{fig:fig8}.

\subsection{Combined result}

Now we combine the constraints on the CKM angle $\alpha$ and strong phase $\delta$
from the experimental measurements studied in this paper.

\begin{figure}[]
\vspace{-40pt}
\begin{minipage}[]{\textwidth}
\centerline{\epsfxsize=\textwidth \epsffile{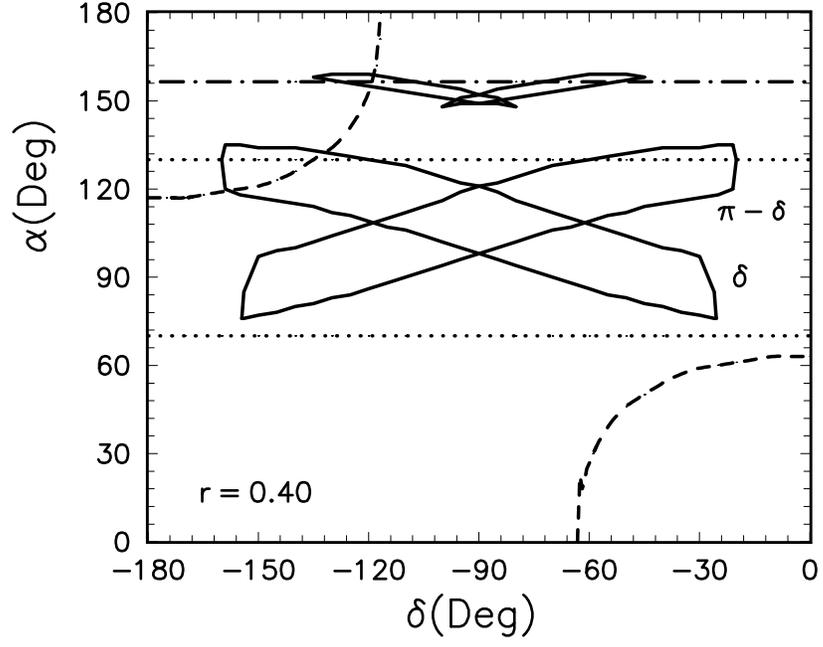}}
\vspace{-20pt}
\caption{The combined constarints on $\alpha$ and $\delta$ for $r=0.4$ and
taking $\spp=0.49 \pm 0.27$, $\app=0.51 \pm 0.19$ as the experimental input.
For details see text.}
\label{fig:fig9}
\end{minipage}
\end{figure}

Fig.\ref{fig:fig9} is the combination of Fig.~5b and Fig.~\ref{fig:fig8} for $r=0.4$.
The upper dot-dashed line shows the direct physical limit of $\alpha \leq 156.4^\circ$ from
the BaBar and Belle's measurements of $\sin{2\beta}$. The regions inside  the solid circles
are allowed by both $\spp^{exp}=-0.49 \pm 0.27$ and $\app^{exp}=0.51 \pm 0.19$
for $r=0.4$. The regions at the upper-left and lower-right corner bounded by short-dashed
curves are allowed by the constraint $\cos{\alpha} \cos{\delta} \geq 0.45$ for $r=0.4$.
The commonly  allowed region can be seen from Fig.~\ref{fig:fig9}, and the
combined constraints on the CKM angle $\alpha$ and the strong phase $\delta$  are
\beq
117^\circ \leq  & \alpha &  \leq 135^\circ, \label{eq:alf1}\\
-160^\circ \leq & \delta &  \leq -132^\circ . \label{eq:alf2}
\eeq
The constraint on $\alpha$ is in very good agreement with that from global
fit, illustrated by the horizontal band between two dots
lines, $70^\circ \lesssim \alpha \lesssim 130^\circ$, and will be improved
along with progress of experimental measurements.
The lower limit of $\alpha \geq 117^\circ$ is much stronger than the limits
obtained before.
The constraint on the strong phase $\delta$, however, depends on
the convention to choose the relative phase between the tree and penguin amplitudes.

From Fig.\ref{fig:fig9}, one can also see that the discrete ambiguity
between $\delta$ and $\pi -\delta$ is resolved by the inclusion of the three measured
branching ratios.

\begin{figure}[]
\vspace{-40pt}
\begin{minipage}[]{\textwidth}
\centerline{\epsfxsize=\textwidth \epsffile{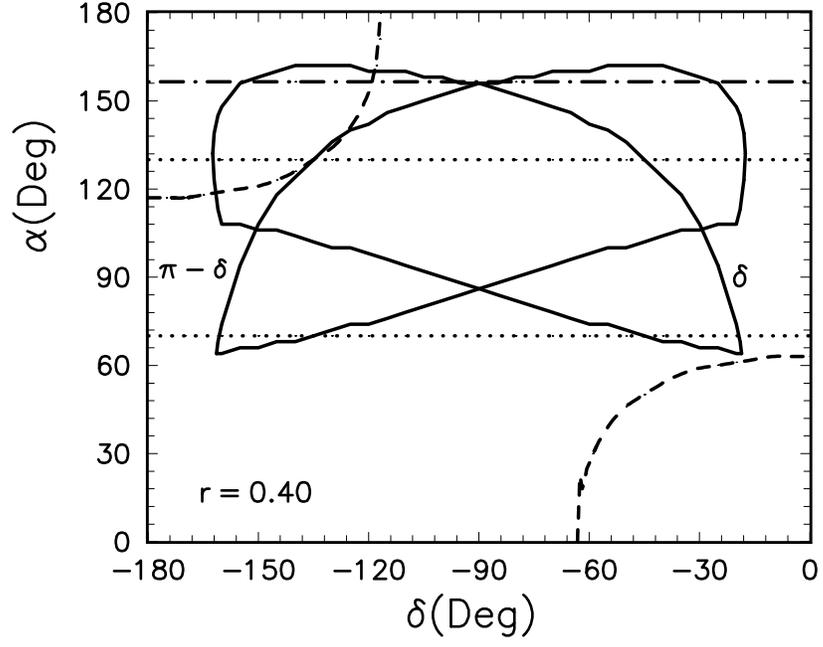}}
\vspace{-20pt}
\caption{The same as Fig.\ref{fig:fig9}, but taking $\spp=0.49 \pm 0.61$, $\app=0.51 \pm 0.23$
as the experimental input. For details see text.}
\label{fig:fig10}
\end{minipage}
\end{figure}

Fig.\ref{fig:fig10} is the combination of Fig.~7b and Fig.~\ref{fig:fig8} for
$r=0.4$. The lower and upper limit on $\alpha$ are dominated by the constraint from
the measured branching ratios and by the direct physical limit $\alpha \leq 156.4^\circ$
respectively, while the constraint on $\alpha$ from the measured
$\spp$ and $\app$  becomes rather weak when one uses the enlarged errors as the $1\sigma$
experimental errors. The allowed regions of the angle $\alpha$ and $\delta$ can then be
read from Fig.~\ref{fig:fig10} directly,
\beq
117^\circ \leq  & \alpha &  \leq 156.4^\circ, \label{eq:alf3}\\
-162^\circ \leq & \delta &  \leq -132^\circ. \label{eq:alf4}
\eeq

The experimental errors of the measured branching ratios of $B \to \pi^+\pi^-, \pi^+ \pi^0$ and
$K^0 \pi^+$ as given in \tab{br} are around $10\%$ now. The errors of the measured $\spp$
and $\app$ as reported by BaBar and Belle Collaborations are still large. Ba comparing the
figures \ref{fig:fig9} and \ref{fig:fig10}, one can see clearly that the improvement of
the experimental measurements of $\spp$ and $\app$ is very important for us to
determine  the CKM angle $\alpha$ reliably.
Besides the error in estimating the magnitude of $|C/T|$,
the main theoretical error in our method may also come from the
neglecting of annihilation type diagrams, which was shown not very
small in some model calculations \cite{luy}.

\section{Summary}

In this paper, we study and try to find constraints
on the CKM angle $\alpha$ and the strong phase $\delta$ from the experimental measurements of
$\ssb$, the CP-violating asymmetries of $B \to \pi^+ \pi^-$ decay, the CP-averaged branching
ratios of $B \to \pi^+ \pi^-, \pi^+ \pi^0$ and $K^0 \pi^+$ decays.

In Section \ref{sec-2}, we give a brief review about the CKM angles and
quote the constraints on these angles obtained from the global fit\cite{ckmfit,fleischer03}.
From the measured $\ssb$\cite{89-201802,66-071102}, the direct physical upper limit on
$\alpha$ is $ \alpha \leq  156.4^\circ$ for $\beta =23.6^\circ$.

In section \ref{sec-3}, we draw constarints on $\alpha$ and $\delta$ from the
measured CP-violating asymmetries of $\spp$ and $\app$. By taking the weighted-average
of BABAR and Belle's measurements as the experimental input, we found the
following constraints
\beq
76^\circ \leq \alpha \leq 135^\circ, \ \ -160^\circ \leq \delta \leq -26^\circ
\eeq
for $\spp=-0.49 \pm 0.27$, $\app = +0.51 \pm 0.19$ and $r=0.3 \pm 0.1$, as shown in
Fig.\ref{fig:fig5}, and
\beq
64^\circ \leq \alpha \leq 162^\circ, \ \ -162^\circ \leq \delta \leq -18^\circ
\eeq
for $\spp=-0.49 \pm 0.61$, $\app = +0.51 \pm 0.23$ and $r=0.3 \pm 0.1$, as shown in
Fig.\ref{fig:fig7}.

In section \ref{sec-4}, we draw constarints on $\alpha$ and $\delta$ from the
measured CP-averaged branching ratios of $B \to \pi^+ \pi^-, \pi^+\pi^0$ and
$B\to K^0 \pi^+$ decays as given in \tab{br}. In the analysis, $SU(2)$ isospin
symmetry and other three assumptions are employed. From the measured decay rates
of $B^+ \to \pi^+ \pi^0$ and $B^+ \to K^0 \pi^+$, one can determine the tree and
penguin amplitudes of $B \to \pi^+ \pi^-$ decay:
$|T|=2.8 \pm 0.3 $, while $|P|=0.80 \pm 0.04$ in GR convention \cite{gronau02}
and $|P|=0.81 \pm 0.19$ in LR convention\cite{luo02}. The destructive interference between
the tree and penguin amplitudes of $B\to \pi^+\pi^-$ decay is then established at $2.6\sigma$
level. The regions
\beq
 63^\circ \leq \alpha \leq 117^\circ, \ \  -117^\circ \leq \delta \leq -63^\circ
\eeq
are excluded by the inequality $\cos{\alpha} \cos{\delta} \geq 0.45$ as given in
Eq.(\ref{eq:cc}), which is a new and important information from the
measurements of the relevant branching ratios.

The combined constraints on the CKM angle $\alpha$ and strong phase $\delta$
have been given in Eqs.(\ref{eq:alf1}-\ref{eq:alf2}), as illustrated in
Figs.\ref{fig:fig9} and \ref{fig:fig10}.
The common allowed regions for $\alpha$ and $\delta$ are
\beq
117^\circ \leq \alpha \leq 135^\circ, \ \
-160^\circ \leq \delta \leq -132^\circ, \label{eq:lf-s2}
\eeq
for $\spp=-0.49 \pm 0.27$, $\app = +0.51 \pm 0.19$ and $r=0.4$,
while
\beq
117^\circ \leq \alpha \leq 156.4^\circ, \ \
-162^\circ \leq \delta \leq -132^\circ, \label{eq:lf-s4}
\eeq
for $\spp=-0.49 \pm 0.61$, $\app = +0.51 \pm 0.23$ and $r=0.4$.

In short, we are able to draw strong constraint on the CKM angle $\alpha$ from currently
available experimental measurements considered in this paper.
The new lower limit on the angle $\alpha$ is dominated by the inequality
$\cos{\alpha} \cos{\delta} \geq 0.45$ and much stronger than those obtained before.
In the SM, the measured value of the angle $\beta$ gives a physical upper
limit on the CKM angle $\alpha$. The measured CP-violating
asymmetries $\spp$ and $\app$ can leads to a strong upper limit on $\alpha$ if we take
$\spp=0.49 \pm 0.27$ and $\app=0.51 \pm 0.19$ as the experimental input.
If we consider the enlarged errors of $\spp$ and $\app$, however, the corresponding upper
limit become weak.
The constraints from different sources are complementary to pin down the allowed region of
$\alpha$. Further improvement of the data will help us to determine $\alpha$ with a
good precision.

\vspace{0.5cm}

\section*{ACKNOWLEDGMENTS}

Z.J.~Xiao and L.B.Guo acknowledges the support by the National Natural Science Foundation
of China under Grants No.~10075013 and 10275035,and by the Research Foundation of
Nanjing Normal University under Grant No.~214080A916. C.D.L\"u acknowledges the support  by
National Science Foundation of China under Grants No.~90103013 and 10135060.

\newpage
\begin{table}
\begin{center}
\caption{Values of the input parameters used in the numerical calculations. Most of them
are quoted from PDG2002 [23]. All masses are in the unit of GeV.}
\label{input}
\begin{tabular}{c c c c c  }
$ |V_{us}|=\lambda $ & $|V_{cb}| $ & $|V_{ub}|$   & $|V_{td}|$ & $V_{tb}$  \\  \hline
$ 0.2196\pm 0.0026 $ & $ (41.2 \pm 2.0)\times 10^{-3} $ & $(3.6 \pm 0.7)\times 10^{-3} $ & $(7.9 \pm 1.5)\times 10^{-3} $
& $ 1$  \\ \hline \hline
$ m_W  $ & $m_t  $ & $m_b^{pole} $     &$m_{B_d} $ &$ m_{B_s}$ \\  \hline
$ 80.42$ & $ 175 $ & $4.80\pm 0.15  $  & $5.279 $ & $5.369 $  \\ \hline \hline
$ f_\pi$ & $f_K $ & $\tau^0$ &$ \tau^+ $ & $\tau^+/\tau^0$ \\  \hline
$0.133$ & $0.158 $ &  $1.542 ps$ & $1.674 ps $ & $1.083\pm 0.017$
\end{tabular} \end{center}
\end{table}

\vspace{2cm}

\begin{table}[t]
\begin{center}
\caption{Experimental measurements of the CP-averaged branching ratios
for $B\to \pi^+ \pi^-, \pi^+ \pi^0$ and $K^0\pi^+$ decays (in units of $10^{-6}$)
as reported by CLEO [24], BaBar [25] and Belle Collaboration [26]. The numbers in last column
are the weighted average. }
\label{br}
\vspace{0.2cm}
\begin{tabular} {lllll}
Channel        & CLEO                  & BaBar & Belle & Average  \\ \hline
$\pi^+ \pi^-$  &$4.5 ^{+1.4 +0.5}_{-1.2 -0.4}$ & $4.7 \pm 0.6 \pm 0.2$          &$4.4 \pm 0.6 \pm 0.3$&$4.6 \pm 0.4$  \\ \hline
$\pi^+ \pi^0$  &$4.6^{+1.8 +0.6}_{-1.6 -0.7} $ & $5.5  ^{+1.0}_{-0.9} \pm 0.6 $ &$5.3 \pm 1.3 \pm 0.5$& $5.3\pm 0.8$  \\ \hline
$ K^0 \pi^+$   &$18.8 ^{+3.7+2.1}_{-3.3-1.8} $ & $17.5 ^{+1.8}_{-1.7} \pm 1.3$  &$22.0 \pm 1.9\pm 1.1$& $19.7 \pm 1.5$ \\
\end{tabular}\end{center}
\end{table}

\newpage

\newpage

\newpage

\newpage

\newpage

\newpage

\newpage

\newpage

\end{document}